\documentclass[prl,aps,twocolumn,superscriptaddress,showpacs]{revtex4-1}

\usepackage{graphicx}
\usepackage{epstopdf}

\begin{document}

\title{Coupled magnetic and ferroelectric excitations in PbFe$_{1/2}$Nb$_{1/2}$O$_{3}$}

\author{C. Stock}
\affiliation{School of Physics and Astronomy, University of Edinburgh, Edinburgh EH9 3JZ, UK}
\author{S.R. Dunsiger}
\affiliation{Physics Department, James Franck Strasse 1, Munich Technical University, D-85748 Garching, Germany}
\author{R. A. Mole}
\affiliation{Forschungsneutronenquelle Heinz Maier-Leibnitz, 85747 Garching, Germany}
\affiliation{ANSTO, Bragg Institute, Locked bag 2001, Kirrawee DC, 2232 NSW, Australia}
\author{X. Li}
\affiliation{Shanghai Institute of Ceramics, Chinese Academy of Sciences, Shanghai, China, 201800}
\author{H. Luo}
\affiliation{Shanghai Institute of Ceramics, Chinese Academy of Sciences, Shanghai, China, 201800}

\date{\today}

\begin{abstract}

A neutron scattering investigation of the magnetoelectric coupling in PbFe$_{1/2}$Nb$_{1/2}$O$_{3}$ (PFN) has been undertaken.   
Ferroelectric order occurs below 400 K, as evidenced by the softening with temperature and subsequent recovery of the zone center transverse optic phonon mode energy ($\hbar \Omega_{0}$).   Over the same temperature range, magnetic correlations become resolution limited on a terahertz energy scale.   In contrast to the behavior of nonmagnetic disordered ferroelectrics (namely Pb(Mg,Zn)$_{1/3}$Nb$_{2/3}$O$_{3}$),  we report the observation of a strong deviation from linearity in the temperature dependence of $(\hbar \Omega_{0})^{2}$.  This deviation is compensated by a corresponding change in the energy scale of the magnetic excitations, as probed through the first moment of the inelastic response.   The coupling between the short-range ferroelectric and antiferromagnetic correlations is consistent with calculations showing that the ferroelectricity is driven by the displacement of the body centered iron site, illustrating the multiferroic nature of magnetic lead based relaxors in the dynamical regime.

\end{abstract}


\maketitle

\textit{Introduction} -- Creating materials with a strong coupling between ferroelectric and magnetic order has been a central goal of research in $3d$ transition metal compounds~\cite{Cheong07:6,Kimura03:426,Scott12:222}.  Fundamentally, this phenomenon is startling, since typically ferroelectricity results from relative 
shifts of negative and positive ions with closed electronic configurations, while magnetism is related to ordering of spins of electrons in incomplete ionic shells.  
While multiferroicity has now been reported in a variety of compounds, as typified by compounds like BiFeO$_{3}$ and BiMnO$_{3}$, coupling between the magnetic and polar orders is often weak and occurs on widely disparate temperature scales~\cite{Santos02:122}.  Magnon-phonon interactions are typically also very weak in improper multiferroics like TbMnO$_{3}$, where only tiny shifts of $\sim$ 0.1 meV in phonon frequencies due to magnetic ordering have been measured~\cite{Rovillain10:81}.  By contrast, we demonstrate a strong intrinsic coupling in a disordered ferroelectric between short range ferroelectric and magnetic orders by measuring the magnetic and lattice fluctuations.  
Recently, the multiferroic nature of disordered systems has attracted considerable attention and the correlation between short range ferroelectric and antiferromagnetic orders has been reported for several different systems~\cite{r1,r2,r3}. For example, in the magnetic relaxor ${(1-x)}$BiFeO$_{3}$-$x$BaTiO$_3$, elastic neutron diffraction studies suggest the polar nanoregions are identical to short range magnetic nanodomains~\cite{r3}.  

Some of the most studied nonmagnetic ferroelectric compounds are the lead based relaxors with the chemical form PbBO$_{3}$, where the B site is a mixture between two different ions~\cite{Park97:82,Ye98:81,Bokov06:41}. PbMg$_{1/3}$Nb$_{2/3}$O$_{3}$ (PMN) and  PbZn$_{1/3}$Nb$_{2/3}$O$_{3}$ (PZN) are prototypical relaxors in which the random occupancy of the B site results in a suppression of a classical ferroelectric phase transition, replaced by short range ferroelectric order.  These polar nanoregions manifest as a strong temperature dependent neutron and x-ray diffuse scattering cross-section which has a significant dynamic component.  The latter component tracks the frequency dependence of the dielectric constant~\cite{Stock10:81,Hirota02:65,Xu04:70,Stock07:76,Gehring09:79}.
A dynamic signature of ferroelectricity in these compounds is a soft and transverse optic phonon mode located at the Brillouin zone center.  The dielectric constant is related to the soft phonon energy ($\hbar \Omega_{0}$) via the Lyndane-Sachs-Teller relation- $1/\epsilon \propto (\hbar \Omega_{0})^{2}$~\cite{Lyddane41:59}.   PbTiO$_{3}$ displays a conventional soft ferroelectric mode, where $(\hbar \Omega)^{2}$ softens to a minimum energy at the structural transition and then recovers linearly below the phase transition~\cite{Shirane70:2,Kempa06:79}.  Even though a long-range ferroelectric ground state is absent in relaxors PMN and PZN, a soft zone center transverse optic mode is present and recovers at the same temperature where static short range polar correlations are onset~\cite{Wakimoto02:65,Stock04:69,Cowley11:60}.

The disordered compound PbFe$_{1/2}$Nb$_{1/2}$O$_{3}$ (PFN) is also based on a perovskite structure with the B site a mixture of nonmagnetic Nb$^{5+}$ and magnetic 
Fe$^{3+}$ ($d^{5}$, $S$=5/2) ions.   Two structural transitions  are reported, the cubic unit cell transforming to tetragonal ($a/c=0.9985$) at 376 K, followed by a 
monoclinic structure ($\beta$=89.94$^{\circ}$) below 355 K~\cite{Lampis99:11}.  The dielectric constant is peaked near 375 K, however is broad in temperature as well as frequency dependent, indicative of short range ferroelectric order~\cite{Majumder06:99}.  A weak cusp at $\sim 150$ K and history dependence below $\sim 20$ K  has been observed in measurements of the bulk magnetization~\cite{bokov,blinc,bhat}. Neutron diffraction measurements have observed short-range three dimensional antiferromagnetic correlations peaked at a wave vector $\vec{Q}$=(1/2,1/2,1/2) coexisting with a magnetic Bragg peak indicative of a long-range magnetically ordered component.  The presence of two distinct lineshapes for the magnetic diffraction has led to models involving two phases defined by different Fe$^{3+}$ clustering sizes~\cite{Kleemann10:105}. 

Motivated by dielectric measurements suggestive of  a coupling between the magnetic and ferroelectric orders (Ref. \onlinecite{Yang04:70})  we present a neutron scattering study of the lattice and magnetic fluctuations in PFN performed at the PUMA spectrometer (FRM2 reactor) on a 1 cm$^{3}$ sample.   We show that the temperature dependent lattice dynamics are dominated by a soft transverse optic mode $(\hbar\Omega)$ measured near the nuclear zone center.   Although  $(\hbar\Omega)^{2}$ recovers, it deviates from the linear response observed in other ferroelectrics (namely PbTiO$_{3}$, PMN, and PZN).  An investigation of the first moment of the magnetic inelastic response suggests that this is compensated by a corresponding change in the energy scale of magnetic excitations, indicative of magnetoelectric coupling {\it in the dynamical regime}.   These measurements are complementary to Raman scattering studies~\cite{correa,garcia}, where the zone center phonon response is also explored, typically at higher characteristic energies.  While Correa {\it et al.}~\cite{correa} report anomalous shifts with sublattice magnetization of the Fe-O phonon mode frequency centered around 87 meV, the shifts are very small ($<3$~cm$^{-1}$ or 0.37 meV).  The so called $F_{2g}$ mode associated with Pb localization around 65 cm$^{-1}$ (8 meV) is completely unaffected by the antiferromagnetic transition observed in their sample.  We show that the magnetoelastic effects observed using inelastic neutron scattering in the low energy regime are much more dramatic. More generally, using Raman techniques, the assignment of phonon modes in the cubic state of the relaxor ferroelectric is controversial~\cite{correa}, making neutron scattering invaluable.

\begin{figure}[t]
\includegraphics[width=8.2cm] {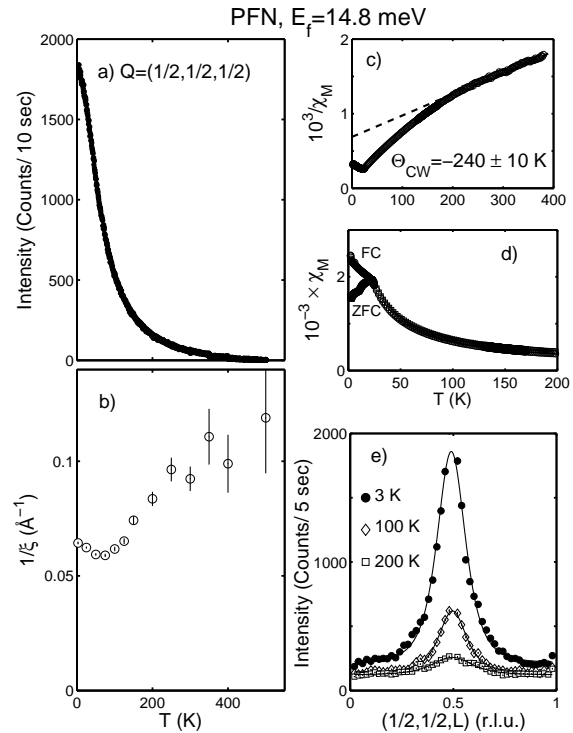}
\caption{\label{mag_elastic} a) the elastic ($\hbar \omega$=0) magnetic intensity as a function of temperature at $\vec{Q}$=(1/2,1/2,1/2).  
b)  the inverse of the correlation length as a function of temperature derived from the constant energy scans displayed in panel e). 
The static magnetization is shown in panels $c)$ and $d)$ under the application of a 100 Gauss field along the [100] axis.}
\end{figure}

$\textit{Static magnetic properties}$ -- Figure \ref{mag_elastic} plots the static magnetic properties of PFN measured through the use of elastic neutron scattering 
(with an energy resolution $\delta E$=1.25 meV=0.30 THz) and static bulk magnetization. Panel $a)$ shows a plot of the elastic magnetic intensity as a function of 
temperature at $\vec{Q}$=(1/2,1/2,1/2), located at the peak of the magnetic intensity (see panel $e$).  It shows only a smooth growth of intensity and no anomaly 
indicative of a well defined phase transition.  The increase in elastic magnetic intensity is mimicked by the static magnetization (panel $c)$ which shows a deviation 
from Curie-Weiss behavior over a similar temperature range.  The combination of the half-integer position in momentum and its broad lineshape indicates the 
origin of the scattering is from clusters of Fe$^{3+}$ ions.  A Lorentzian squared lineshape was used to model the data, motivated by random fields~\cite{Birgeneau83:28}.
Such random fields are thought to be the underlying cause of the avoided long range magnetic ordering, which would be characterized by a Bragg peak.  
Similar ideas have been proposed to understand the ferroelectric properties of disordered PMN and PZN~\cite{Westphal92:68,Fisch03:67}.  
Given the large vertical resolution on the PUMA spectrometer (Ref. \onlinecite{Link00:276}), we integrated over the vertical direction analytically, giving 
the ${3/2}$ power to the momentum dependence shown below.    The magnetic correlation lengths (panel $b$) were thus extracted from fits to a resolution convolution of the following lineshape (examples illustrated in $e$):
\begin{eqnarray}
I(\vec{Q})=C {(\gamma r_{0})^2 \over 4} g^{2} f^2(Q) m^{2} e^{-\langle u^{2} \rangle Q^{2}} {V^{*} {\xi^{3}/\pi^{2}} \over [1+(|\vec{Q}-\vec{Q}_{0}|\xi)^{2}]^{3/2}}, 
\nonumber
\end{eqnarray}
where $C$ is the calibration constant, $(\gamma r_{0})^{2}/4$ is 73 mbarns sr$^{-1}$, $g=2$ is the Land{\'e} factor, $f^{2}(Q)$ is the Fe$^{3+}$ magnetic form factor, $m$ is the magnetic moment size, $V^{*}$ is the volume of the Brillouin zone, $\xi$ the correlation length, $e^{-\langle u^{2} \rangle Q^{2}}$ is the Debye-Waller factor, and $a$ is the lattice constant.    Performing scans along all high symmetry directions ([111], [110], and [001]), the magnetic correlation length is found to be spatially isotropic within error.   As demonstrated in Fig. \ref{mag_elastic} $b)$, this quantity never diverges, but saturates at the small value of $\xi$=17\AA\ with an inflection point around $\sim$ 80 K.
The inflection point in the temperature dependent correlation length is mimicked by the static magnetization 
presented in Fig. \ref{mag_elastic} $d)$, albeit at much lower temperatures of $\sim$ 25 K.  The divergence between field-cooled (FC) and zero-field cooled (ZFC) magnetization reflects the development of a spin-glass-like state at low temperatures and is close to the $\sim$ 20 K anomaly observed using muon spin relaxation~\cite{Rotaru09:79}.  While there is apparent sample dependence evidenced by the different values quoted in the literature for this divergence in the magnetization (27.6 K~\cite{Kumar08:93},$\sim$20 K~\cite{Rotaru09:79}, and 10.6 K~\cite{Kleemann10:105}),  as pointed out in Ref. \onlinecite{Falqui05:109}, the exact value may be strongly technique and field dependent.  The differing onset temperatures of spin glass-like ordering produced by different techniques (neutrons - 80 K compared to static magnetization - 25 K) which probe different timescales is similar to results published on canonical spin glasses and frustrated magnets, where the magnetic correlations are dominated by slow fluctuations over a broad frequency range, sampled with differing energy resolutions~\cite{Stock10:105,Murani78:41}. 


The high temperature fit of the static magnetization yields a Curie constant of $k_{B}\Theta_{CW}$=-240 $\pm$ 10 K which is a measure of the 
average Fe$^{3+}$-Fe$^{3+}$ exchange coupling.  The negative sign indicates  antiferromagnetic coupling between the spins.  Within the mean-field 
approximation $k_{B}\Theta_{CW}=-{1\over3} z J S(S+1)$ (with $z$=3 neighbors and $S={5\over2}$) gives an estimate of the average $J=2.4 \pm 0.4$ meV $\sim$ 28 K.  
This implies a magnetic band width of the order of 10 meV ($\sim 2SJ$). The energy scale of the magnetic coupling is consistent with the presence of strong low energy magnetic spectral weight at high temperatures, within the spectrometer resolution (Fig. \ref{mag_elastic}$e$).
%
%
\begin{figure}[t]
\includegraphics[width=8.2cm] {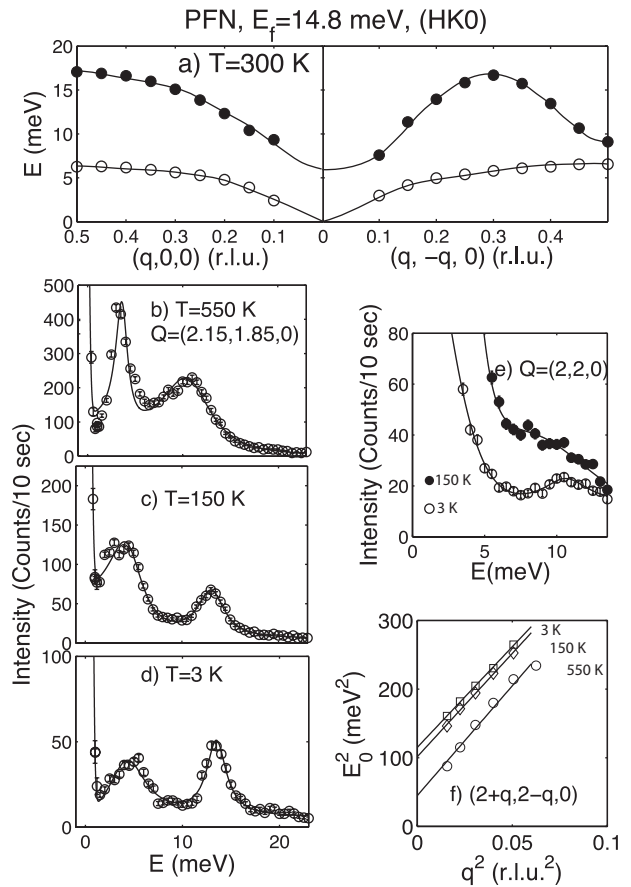}
\caption{\label{phonon} a) the phonon dispersion for both the T$_{1}$ and T$_{2}$ modes.  b)-d) representative constant-Q scans taken at $\vec{Q}$=(2.15, 1.85, 0)  e) constant-Q scans at the zone center at 150 K (filled circles) and 3 K (open circles) f) the frequency squared of the soft transverse optic mode as a function of wavevector squared near the zone center.}
\end{figure}

$\textit{Ferroelectric properties}$ -- The  lattice dynamics are similar in structure and energy scale to 
PMN and PbTiO$_{3}$~\cite{Swainson09:79,Tomeno06:73,Hlinka06:73}.   The phonons are described by a transverse optic mode which is gapped at the nuclear zone center 
and a lower energy acoustic mode which is gapless.   The dispersion near the nuclear zone center is shown in Fig. \ref{phonon} $a)$ and illustrative 
constant-$\vec{Q}$ scans are shown in panels $b-d)$.   The constant-$Q$ scans show there is little change in frequency in the low-energy acoustic mode.  
However, the higher energy optic mode gradually hardens as the temperature is decreased, as expected for the recovery from a structural transition.  
In a similar manner to the case of PMN, we observed the optic mode is over damped in energy near the nuclear zone center, plotted in panel $e)$, where a very broad 
and unresolvable peak was observed at 150 K.  A more well defined peak at $\sim$ 11 meV is observed at 3 K~\cite{Stock12:86}.    To extract the zone center soft mode energy, we rely on the optic mode frequencies at finite $q$ away from the zone 
center and fit the results to $(\hbar \Omega(q))^{2}=(\hbar \Omega_{\circ})^{2}+\alpha q^{2}$. $\Omega_{\circ}$ and $\alpha$  are respectively the optic mode 
frequency at the zone center and a temperature independent measure of the curvature near the zone center.   Representative results from this analysis are plotted in 
panel $f)$ and where the zone center frequency could be measured, good agreement was observed.  This method has been applied before to the relaxors and found to be in good agreement with zone center scans, as well as Raman data~\cite{Cao08:78,Shirane70:2}.

\begin{figure}[t]
\includegraphics[width=9.0cm] {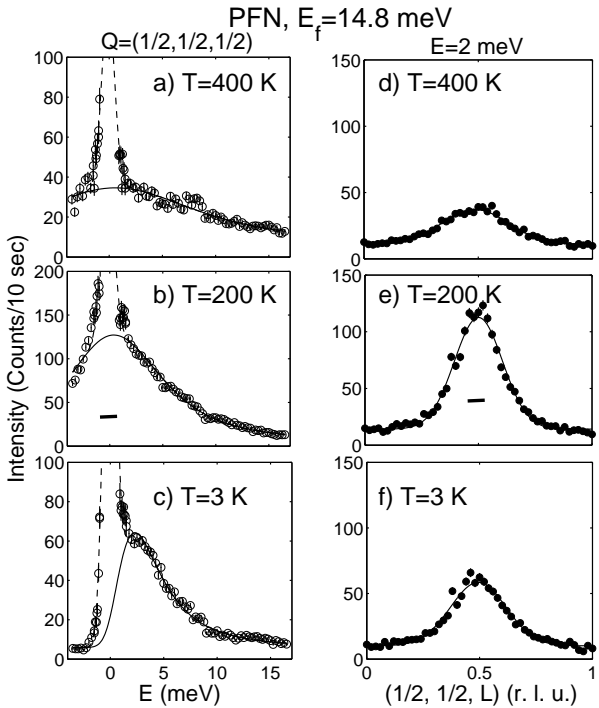}
\caption{\label{mag_inelastic} $a-c)$ are constant-$\vec{Q}$=(1/2,1/2,1/2) scans at several temperatures.  The solid curves represent a fit to the simple harmonic oscillator described in the text.  $d-f)$ are constant $\hbar \omega$=2 meV scans at the sample temperatures.  The resolution full-widths are represented by the solid bars.}
\end{figure}

$\textit{Magnetic dynamics}$ -- The temperature dependent magnetic dynamics are illustrated in Fig. \ref{mag_inelastic} through constant $\vec{Q}$=(1/2,1/2,1/2) 
scans (panels $a-c$) and constant $\hbar\omega$= 2 meV scans (panels $d-f$).  The fluctuations are dominated by a peak at $\vec{Q}$=(1/2, 1/2, 1/2) which is both 
momentum and energy broadened, characteristic of short range and glass-like dynamics.  The dynamic magnetic response is not dispersive and is overdamped for all temperatures investigated.  Constant $\vec{Q}$=(1/2,1/2,1/2) scans were fit to a resolution convolved damped harmonic oscillator as described in the supplementary information.
%
%
%
%
\begin{figure}[t]
\includegraphics[width=8.0cm] {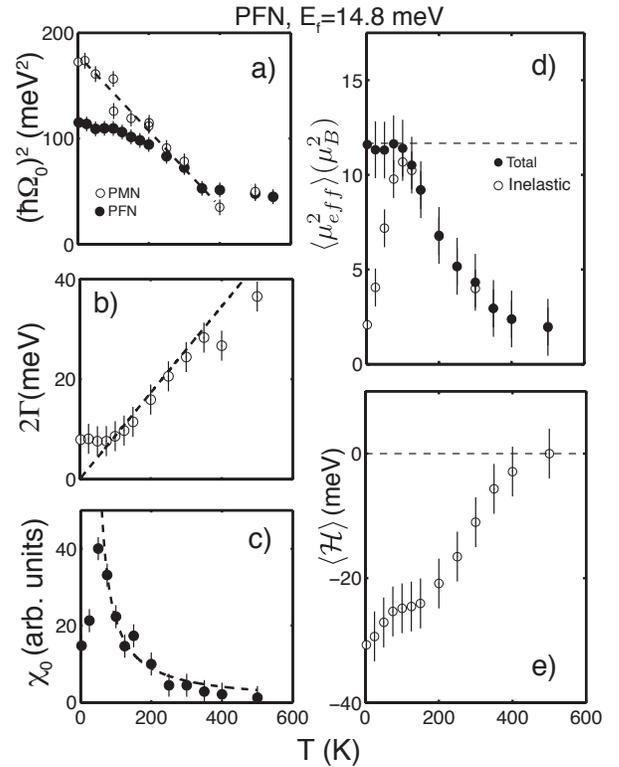}
\caption{\label{param} $a)$ the soft transverse optic phonon frequency squared as a function of temperature for both PFN and PMN. $b)$ and $c)$ illustrates the magnetic fitting parameters described in the text.  $b)$ shows the line width as a function of temperature and $c)$ the susceptibility.   $d)$ the zeroeth moment for the static and dynamic components on the THz timescale. $e)$ shows the change in the first moment with temperature.}
\end{figure}

The temperature variation of $(\hbar\Omega_{0})^{2}$ (proportional to $1/\epsilon$), the timescale of the magnetic fluctuations ($2\Gamma \sim 1/\tau$), and the 
susceptibility ($\chi_{0}$) at $\vec{Q}$=(1/2, 1/2, 1/2) are illustrated in Fig. \ref{param} $a-c)$.   The energy of the soft optic mode in PMN is also 
plotted in Fig. \ref{param} $a)$, showing a linear recovery down to low temperatures~\cite{Stock05:74,Wakimoto02:65}.  For comparison, the data from PMN has been scaled by a factor of 1.4 to agree with PFN at 400 K.  While at temperatures above $\sim$ 100 K the zone center energy tracks the measured response in prototypical relaxors (such as PMN), a significant deviation from the linear recovery is observed at low temperatures.   

The linewidth of the magnetic fluctuations measured at $\vec{Q}$=(1/2,1/2,1/2) is plotted in Fig. \ref{param} $b)$ and illustrates a linear decrease towards saturation at $\sim$ 80 K.   The dashed line is given by $2\Gamma=k_{B}T$, demonstrating that the energy scale of the magnetic fluctuations at high temperatures is set by $k_{B}T$.  Figure \ref{param} $c)$ illustrates the susceptibility $\chi_{0}$ derived from fits shown in Fig. \ref{mag_inelastic}. The dashed line is a high temperature fit to $1/(T-\Lambda)$ with a deviation at $\sim$ 80 K - the same temperature as the inflection point in the elastic correlation length (Fig. \ref{mag_elastic} $b$) and the saturation of the inelastic linewidth (Fig. \ref{param} $b$). 

Given the broad nature of the magnetic response in momentum and energy of our data and the ambiguity over its functional form, it is important to ensure all the spectral weight is accounted for.  In general, the total integrated magnetic intensity over all momentum and energy transfers is a conserved quantity satisfying 
the zeroeth moment sum rule. We calculated the zeroeth moment sum (Fig. \ref{param} $d$) by integrating the magnetic intensity over the $Q$ and $E$ range presented in Fig.\ref{mag_inelastic}.  The data were normalized against the known cross section of an acoustic phonon.   The dashed line is the 
expected value based on a $S={5\over2}$ moment for Fe$^{3+}$, indicating the experiment indeed probed all of the magnetic scattering. The filled points are the 
total spectral weight summing over the elastic and inelastic channels, while the open circles represent the inelastic component.  The difference at low temperatures 
corresponds to an estimate of the total spectral weight measured to be static, or within resolution limits (0.3 THz).  The decrease of the spectral weight with 
increasing temperature indicates the magnetic response extends to higher energies not directly probed.

The deviation of $(\hbar\Omega)^{2}$ from linearity (Fig. \ref{param}$a$) differs from nonmagnetic counterparts PMN and PZN and is suggestive of a coupling to another energy scale.  Given the magnetic response is strongly damped (Fig. \ref{mag_inelastic} ), we have investigated the magnetic energy scale 
through a study of the first moment $\langle \mathcal{H} \rangle$ calculated from the Hohenberg Brinkman (Ref. \onlinecite{Hohenberg74:10}) sum rule.  The sum rule is a general result for isotropic correlations and is applicable to the case where no sharp peak is observed in a constant $Q$ scan:
\begin{eqnarray}
\label{first_moment} \langle \mathcal{H} \rangle=-{3\over4}{{\int_{-\infty}^{\infty}dE (E\chi''(Q,E))}\over{(1-\cos(\vec{Q}\cdot\vec{d}))}}.
\nonumber
\end{eqnarray}
The integration was performed at $\vec{Q}$=(1/2,1/2,1/2) and $d$ is the distance between nearest neighbors.   The change in the first moment with temperature is plotted in Fig. \ref{param} $e)$, showing a substantial reduction with decreasing temperature.  The change is of order the expected change in energy between the soft optic phonon in PFN and PMN, indicating a coupling between the two orders.

$\textit{Summary and Conclusions}$ -- Our results illustrate that in the presence of short-range magnetic order, the soft phonon dynamics are directly altered from 
the linear 
recovery in $(\hbar \Omega)^{2}$ observed in classic and disordered ferroelectrics.  The first moment, a measure of the magnetic energy scale, illustrates the 
change in energy is taken up by the magnetic spin terms demonstrating coupling between the two orders.    
Such a coupling may be expected given the local bonding environment in PFN.  The ferroelectric order in compounds of the form ABO$_{3}$ has been found to be 
determined by the condensation of predominately the Last and Slater phonon modes, with significant contributions from shifts in the A and O sites~\cite{Harada70:A26}.  
In the fully ordered case of PFN, the exchange interaction between two Fe$^{3+}$ ions would involve orbitals from Pb and also O.  Therefore, the hybridization 
associated with ferroelectric order would be expected to alter the exchange pathways coupling magnetic ions.  Such a scenario has been suggested to exist in fully 
site ordered EuTiO$_{3}$~\cite{Katsufuji01:64}.  An alternate explanation is proposed as a result of calculations (Ref. \onlinecite{Wang01:288}) which suggest that 
the ferroelectricity in PFN predominately originates from the displacement of the Fe$^{3+}$ site, the ratio of the displacement of Fe$^{3+}$ to Nb$^{5+}$ being greater 
than 10.  Such a large difference in 
displacement could provide a route for explaining the strong coupling between the two orders observed here.

Strong magnetoelectric coupling appears to be favored in disordered systems where the symmetry contraints of the lattice are relaxed, as in compounds like 
pervoskite (Sr,Mn)TiO$_{3}$ and nonperovskite (Ni,Mn)TiO$_{3}$~\cite{r1,r2}.  However, enhanced dielectric constants have been reported 
previously in a number of candidate materials for multiferroicity, which were later shown to arise from nanoscale disorder~\cite{zhu07,ruff12}. Using a microscopic 
technique robust against such extrinsic effects, we have found evidence of coupling between the short range magnetic and ferroelectric orders in the archetypal 
ferroelectric PFN.  The temperature dependence of $(\hbar \Omega_{0})^{2}$ associated with the soft transverse optic mode, sensitive to ferroelectric correlations, 
deviates strongly from the linear recovery observed in classic ferroelectrics as well as prototypical nonmagnetic relaxors Pb(Mg,Zn)$_{1/3}$Nb$_{2/3}$O$_{3}$.  

We are grateful for funding from EU-NMI3, the Carnegie Trust for the Universities of Scotland, STFC, and Deutsche Forschungsgemeinschaft Grant TRR 80.

\section{Appendix} 
Here we present supplementary information regarding the experimental details, spectrometer calibration, and sum rules used in the main text.  
Supplementary data regarding the momentum dependence of the magnetic scattering are also presented.  The data demonstrate the absence of well-defined spin waves and 
show that the magnetic excitations are represented by strongly overdamped fluctuations characteristic of the short range magnetic order. 

\subsection{Experimental details for the neutron scattering and susceptibility measurements}

Neutron scattering measurements were performed on the PUMA thermal triple-axis spectrometer located at the FRM2 reactor (Garching, Germany).  
Two sets of measurements were performed with the 1 cm$^{3}$ (with lattice constant $a$=4.01 \AA)  sample to measure both the magnetic and lattice fluctuations.  
To measure the phonon dispersion curves and temperature dependence, the sample was oriented in the (HK0) scattering plane.  The 
magnetic scattering was investigated with the sample mounted in the (HHL) plane.    In both sets of measurements, the sample was cooled in a closed cycle refrigerator.  
A PG(002) vertically focused monochromator was used to select an incident energy E$_{i}$ and the final energy was fixed at E$_{f}$=14.8 meV using a PG(002) flat 
analyzer crystal.  The energy transfer was then defined as $\hbar \omega$=E$_{i}$-E$_{f}$. To measure the phonon curves, it was desirable to obtain a high count 
rate at the expense of momentum resolution and therefore horizontal focusing was used on both the monochromator and the analyser.  
The horizontal focusing, on both the incident and scattered sides was removed for studies of the static and fluctuating magnetic response.  
Higher order contamination was reduced through the use of a pyrolytic graphite filter in the scattered beam.  
The counting time was determined by a low efficiency monitor placed in the incident beam and was corrected for variable contamination by higher order scattering from the monochromator using the same calibration described elsewhere.~\cite{Stock04:69} 

Magnetization measurements were performed using a Quantum Design Materials Properties Measurement System (MPMS) on a small 7.0 mg piece of PFN taken from the same crystal growth batch.  A field of 100 Gauss was applied along the $a$ axis and measurements under field cooled (FC) and zero-field cooled conditions were performed.

\subsection{Spectrometer calibration constant derived from acoustic phonons}

To calculate the zeroeth and first moments of the magnetic scattering, we have put the magnetic intensities on an absolute scale.  
The calibration constant for the experiment was obtained by measuring a low-energy acoustic phonon.  In the long-wavelength (low $q$) limit, it can be assumed 
that we are in the hydrodynamic regime where only the center of mass is moving and the structure factor for the acoustic phonon will match that of the nearby 
nuclear Bragg peak.   In the setup used on PUMA with an incident beam monitor with an efficiency $\propto {1\over{\sqrt{E}}}$, the measured energy integrated 
intensity takes the form
\begin{eqnarray}
I(\vec{Q})=A\left({\hbar \over {2 \omega_{0}} } \right) [1+n(\omega_{0})]|F_{N}|^{2} {{Q^{2} \cos^{2}(\beta)} \over {M}}e^{-2W}.
\end{eqnarray}
where $A$ is the spectrometer calibration constant, $\hbar \omega_{0}$ is the acoustic phonon frequency, $[1+n(\omega_{0})]$ is the Bose factor, $|F_{N}|^{2}$ the 
structure factor of the nuclear Bragg peak, $M$ the mass of the unit cell, and $e^{-2W} \sim 1$ is the Debye Waller factor.  For the measured magnetic scattering 
in the (HHL) scattering plane, we have used an acoustic phonon measured at $\vec{Q}$=(0.15,0.15,2) and T=300 K as a reference with a horizontally flat monochromator and analyzer.

\subsection{Lineshape- inelastic magnetic response}

To describe the broad overdamped lineshape characterizing the magnetic dynamics, a relaxational form determined by a single energy scale $\Gamma\propto1/\tau$ described 
by $\chi''(Q,E)\propto \chi_{0}(Q)E\Gamma/(E^{2}+\Gamma^{2})$ was initially fit to the constant $Q$ scans shown in Fig. 3 of the main text.  
Using this form for the magnetic dynamics, $\chi_{0}$ is related to the real part of the susceptibility.  While this line shape described the data well at 
high temperatures, it failed to fit the response below $\sim$ 100 K.  To correct this, following Ref. \onlinecite{Shapiro88:xx}, we fit the following damped 
harmonic oscillator lineshape to all temperatures.
\begin{eqnarray}
\label{SHO} S(E)= {\chi_{0}}  [n(E)+1] \times \\
\left( {{1}\over {[1+
{{\left(E- E_{0}\right)^{2}}\over {\Gamma^{2}}} }]} - {{1}\over {[1+
{{\left(E+ E_{0}\right)^{2}}\over {\Gamma^{2}}} }]} \right),
\nonumber
\end{eqnarray}
where $\chi_{0}$ is a measure of the strength of the magnetic scattering, $[n(E)+1]$ is the thermal population (or Bose) factor, $\hbar \Omega$ the mode position, and $\Gamma$ is the half-width.  To account for elastic scattering from static (defined by our resolution width) correlations, we have included a Gaussian in the fit centered at the elastic energy position.  
The parameter E$_{0}$ was found to be temperature independent with E$_{0}$=0.5 $\pm$ 0.2 meV and can be physically interpreted as a magnetic anisotropy energy scale.

\begin{figure}[t]
\includegraphics[width=8.2cm] {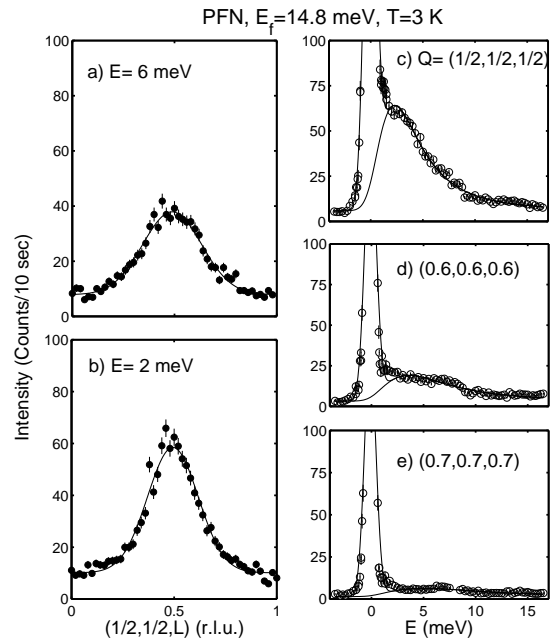}
\caption{\label{dispersion} The momentum dependence of the scattering at T=3 K.  $a-b)$ illustrate constant energy scans at E=6 and 2 meV.  $c-e)$ show constant-Q scans taken at positions close to $\vec{Q}$=(${1\over 2}, {1\over 2}, {1\over 2}$)}
\end{figure}

\subsection{Zeroeth moment sum rule for magnetic scattering}

The total integrated magnetic intensity over all momentum and energy transfers is a conserved quantity satisfying the zeroeth moment sum rule.  Accounting for the orientation factor in magnetic neutron scattering and the fact that there is ${1 \over 2}$ a Fe$^{3+}$ site per unit cell, the integral over $S(\vec{Q},E)$ is,
\begin{eqnarray}
\langle \mu_{eff}^{2} \rangle = \int dE \int d^{3}Q S(\vec{Q},E)=...\nonumber \\
{1\over \pi} \int dE \int d^{3}Q [1+n(E)]\chi''(\vec{Q},E)=... \nonumber \\
{2 \over 3} g^{2} \mu_{B}^{2} S(S+1) \times {1 \over 2}.
\end{eqnarray}
Setting $S={5 \over 2}$ gives a total expected integral of 11.7 $\mu_{B}^{2}$.  This is in agreement with the total integrated moment discussed in the main text. \subsection{$\vec{Q}$-E dependence}

The zeroeth and first moment analysis outlined in the main text relies on a knowledge of the momentum dependence of the magnetic scattering with energy transfer.   Fig. \ref{dispersion} illustrates the momentum dependence of the magnetic scattering through both constant energy (panels $a-b)$ and also constant-Q scans (panels $c-e)$).  The constant energy scans have been fitted to a Gaussian centered at $\vec{Q}$=(${1\over 2}, {1\over 2}, {1\over 2}$) and show only a single central peak.  The constant-Q scans also do not display 
any sign of spin waves or dispersion of the magnetic excitations.  The first moment analysis has used the fact that the excitations are peaked only near $\vec{Q}$=(${1\over 2}, {1\over 2}, {1\over 2}$) and this is substantiated by the results.


%

\end{document}